\documentclass[useAMS,usegraphicx,usenatbib]{mn2e}

\usepackage{times}

\newcommand{\Msun}{\ensuremath{\,{\rm M}_\odot}}                  % Solar mass symbol
\newcommand{\Rsun}{\ensuremath{\,{\rm R}_\odot}}                  % Solar radius symbol
\newcommand{\Teff}{\ensuremath{T_{\rm eff}}}                      % Effective temperature symbol
\newcommand{\Mjup}{\ensuremath{\,{\rm M}_{\rm Jup}}}              % Jupiter mass symbol
\newcommand{\Rjup}{\ensuremath{\,{\rm R}_{\rm Jup}}}              % Jupiter radius symbol
\newcommand{\Teq}{\ensuremath{T_{\rm eq}^{\,\prime}}}             % Equilibrium temperature symbol
\newcommand{\safronov}{\ensuremath{\Theta}}                       % Safronov number symbol
\newcommand{\mss}{\,m\,s$^{-2}$}                                  % m/s^2 symbol
\newcommand{\as}{\ensuremath{^{\prime\prime}}}                    % Arcsecond symbol
\newcommand{\FeH}{\ensuremath{\left[\frac{\rm Fe}{\rm H}\right]}} % [Fe/H] symbol
\newcommand{\pjup}{\ensuremath{\,\rho_{\rm Jup}}}                 % Jupiter density symbol
\newcommand{\psun}{\ensuremath{\,\rho_\odot}}                     % Solar density symbol
\newcommand{\mc}[1]{\multicolumn{2}{c}{#1}}
\newcommand{\mcc}[1]{\multicolumn{3}{c}{#1}}
\newcommand{\er}[3]{\ensuremath{#1^{+#2}_{-#3}}}
\newcommand{\erc}[3]{\mc{\ensuremath{#1^{+#2}_{-#3}}}}
\newcommand{\ercbf}[3]{\mc{\ensuremath{\mathbf{#1}^\mathbf{+#2}_\mathbf{-#3}}}}

\newcommand{\ercc}[3]{\mcc{\ensuremath{#1^{+#2}_{-#3}}}}
\newcommand{\ermcc}[5]{\mcc{\ensuremath{{#1\,^{+#2}_{-#3}}\,^{+#4}_{-#5}}}}
\newcommand{\reff}[1]{{#1}}

\title[The ultra-short period planet WASP-103\,b]
      {Contamination from a nearby star cannot explain the anomalous transmission spectrum of the ultra-short period giant planet WASP-103\,b}

\author[Southworth \& Evans]
       {John Southworth and Daniel F.\ Evans \\ Astrophysics Group, Keele University, Staffordshire, ST5 5BG, UK}

\begin{document} \maketitle %%%%%%%%%%%%%%%%%%%%%%%%%%%%%%%%%%%%%%%%%%%%%%%%%%%%%%%%%%%%%%%%%%%%%%%%%%%%%%%%%%%%%%%%%%%%%%%%%%%%%%%%%%%%%%%%%%%%%%%%%
%%%%%%%%%%%%%%%%%%%%%%%%%%%%%%%%%%%%%%%%%%%%%%%%%%%%%%%%%%%%%%%%%%%%%%%%%%%%%%%%%%%%%%%%%%%%%%%%%%%%%%%%%%%%%%%%%%%%%%%%%%%%%%%%%%%%%%%%%%%%%%%%%%%%%

\begin{abstract}
The planet in the WASP-103 system is an excellent candidate for transmission spectroscopy because of its large radius and high temperature. Application of this technique found a variation of radius with wavelength which was far too strong to be explained by scattering processes in the planetary atmosphere. A faint nearby star was subsequently detected, whose contamination of the transit light curves might explain this anomaly. We present a reanalysis of published data in order to characterise the faint star and assess its effect on the measured transmission spectrum. The faint star \reff{has a mass of $0.72 \pm 0.08$\Msun} and is almost certainly gravitationally bound to the planetary system. We find that its effect on the measured physical properties of the planet and host star is small, amounting to a planetary radius larger by $0.6\sigma$ and planetary density smaller by $0.8\sigma$. Its influence on the measured transmission spectrum is much greater: the spectrum now has a minimum around 760\,nm and opacity rises to both bluer and redder wavelengths. It is a poor match to theoretical spectra and the spectral slope remains too strong for Rayleigh scattering. The existence of the faint nearby star cannot therefore explain the measured spectral properties of this hot and inflated planet. We advocate further observations of the system, both with high spatial resolution in order to improve the measured properties of the faint star, and with higher spectral resolution to confirm the anomalous transmission spectrum of the planet.
\end{abstract}


%%%%%%%%%%%%%%%%%%%%%%%%%%%%%%%%%%%%%%%%%%%%%%%%%%%%%%%%%%%%%%%%%%%%%%%%%%%%%%%%%%%%%%%%%%%%%%%%%%%%%%%%%%%%%%%%%%%%%%%%%%%%%%%%%%%%%%%%%%%%%%%%%%%%%

\section{Introduction}                                                                                                              \label{sec:intro}

Hot Jupiters were the first type of extrasolar planet to be discovered, for both the radial velocity and transit methods \citep{MayorQueloz95nat,Henry+00apj,Charbonneau+00apj}, their detection being aided by their comparatively large radii and short orbital periods. They were also the first extrasolar planets whose atmospheres were detected \citep{Charbonneau+02apj,Vidal+04apj}, helped by their often-large atmospheric scale heights. At this point, approximately 30 transiting hot Jupiters have been studied using the method of transmission spectroscopy, where opacity in the planetary atmosphere is probed by measuring the size of the planet as a function of wavelength \citep[e.g.][]{Sing+16nat}.

Such analyses can also be performed using transmission photometry, where wavelength resolution is achieved by using multiple passbands rather than via a spectroscopic approach \citep[e.g.][]{Mallonn+15aa2}. Versus transmission spectroscopy, the method of transmission photometry typically requires more observing time and has a lower wavelength resolution, but can be performed on smaller telescopes and is less subject to systematic errors due to Earth's atmosphere and instrumental effects \citep{Me+12mn2}. Rayleigh scattering has so far been detected in four transiting hot Jupiters using transmission photometry: GJ\,3470\,b \citep{Nascimbeni+13aa2,Biddle+14mn,Dragomir+15apj}, WASP-103 \,b \citep[][hereafter Paper\,I]{Me+15mn}, and tentatively in GJ\,1214\,b \citep{Demooij+12aa} and Qatar-2\,b \citep{Mancini+14mn}.

The WASP-103 system \citep[][hereafter G14]{Gillon+14aa} stands out in this list as having the hottest star, the largest planet, and a highly significant detection of the Rayleigh scattering signal (7.3$\sigma$) which is, however, much stronger than expected. Adopting the MassSpec concept from \citet{DewitSeager13sci} leads to a measurement of the planetary mass which is a factor of five smaller than the dynamical mass measurement (Paper\,I). Since this work, a faint and cool nearby star has been detected with a very small sky-projected separation from the WASP-103 system \citep{WollertBrandner15aa}. The purpose of the current work is to revisit the analysis of WASP-103 to determine the effect of the presence of this faint companion star on the measured properties of the system, and see if it can provide an explanation for the anomalous transmission spectrum of the planet. The presence of a faint nearby star has previously been shown to affect both transmission \citep{Lendl+16aa} and emission \citep{Crossfield+12apj} spectroscopy.

%%%%%%%%%%%%%%%%%%%%%%%%%%%%%%%%%%%%%%%%%%%%%%%%%%%%%%%%%%%%%%%%%%%%%%%%%%%%%%%%%%%%%%%%%%%%%%%%%%%%%%%%%%%%%%%%%%%%%%%%%%%%%%%%%%%%%%%%%%%%%%%%%%%%%

\section{Reanalysis of the light curves}                                                                                               \label{sec:lc}

In Paper\,I we presented light curves of 11 transits of WASP-103 obtained using three telescopes and seven optical passbands. Eight transits were observed using the 1.54\,m Danish Telescope at ESO La Silla, seven through a Bessell $R$ filter and one through a Bessell $I$ filter. Two transits were observed in four passbands simultaneously (similar to the Gunn $griz$ bands) using the GROND imager \citep{Greiner+08pasp} on the MPG 2.2\,m at the same site. The final transit was observed using the 2.15\,m telescope at CASLEO, Argentina, and will not be considered further in this work because of its significantly greater scatter and more complex continuum normalisation (Paper\,I).

\subsection{Accounting for the companion star}

\begin{table*} \centering \caption{\label{tab:lr} \reff{The fractional contribution of the faint nearby
star to the total light of the WASP-103 system, assessed using the two sources of magnitude differences.}}
\begin{tabular}{l c c}
\hline
Parameter & Value from $\Delta i$ and $\Delta z$ & \reff{Value from $\Delta J$ and $\Delta K_s$} \\
\hline
${\Teff}_{\rm,comp}$ (K)                  & \er{3377}{743}{199} & \er{4405}{85}{80}   \\
Fractional contribution in Bessell $R$    & $0.023  \pm 0.023 $ & $0.0525 \pm 0.0040$ \\
Fractional contribution in Bessell $I$    & $0.055  \pm 0.017 $ & $0.0673 \pm 0.0040$ \\
Fractional contribution in GROND $g$      & $0.0068 \pm 0.0068$ & $0.0246 \pm 0.0032$ \\
Fractional contribution in GROND $r$      & $0.019  \pm 0.019 $ & $0.0492 \pm 0.0040$ \\
Fractional contribution in GROND $i$      & $0.042  \pm 0.020 $ & $0.0641 \pm 0.0041$ \\
Fractional contribution in GROND $z$      & $0.089  \pm 0.028 $ & $0.0766 \pm 0.0037$ \\
\hline \end{tabular} \end{table*}

In Paper\,I we presented a high-resolution image of the sky area surrounding WASP-103 using the {\it lucky imaging} technique and the Two Colour Imager also on the Danish 1.54\,m telescope \citep{Skottfelt+15aa}. The image had a FWHM of 5.9 pixels (0.53\,arcsec) in both spatial scales and showed no evidence for stars sufficiently close by to contaminate the light curves obtained of this object. The lower limit on the spatial resolution of this instrument is approximately 0.5\,arcsec, imposed by triangular coma present in the telescope optics (\citealt{Skottfelt+15aa}; see also \citealt{Evans+16aa}).

Subsequent to this work, \citet{WollertBrandner15aa} presented the discovery of a companion star to the WASP-103 system, based on observations using the AstraLux lucky imager at Calar Alto Observatory, Spain. The star is separated by $0.242 \pm 0.016$\,arcsec at a position angle of $132.66 \pm 2.74^\circ$, \reff{so was too close to be apparent on our own lucky imaging (Paper\,I). It is fainter by $\Delta i = 3.11 \pm 0.46$ and $\Delta z = 2.59 \pm 0.35$ mag, so is likely to be significantly cooler than WASP\,103\,A and therefore} impose a wavelength-dependent contamination on photometry of the planetary system.

The two magnitude differences were used to determine the fraction of contaminating light in the passbands used in Paper\,I following the method outlined by \citet{Me10mn} and \citet{Me+10mn}. In brief, theoretical spectra from {\sc atlas9} model atmospheres and the known effective temperature of the planet host star ($\Teff = 6110 \pm 160$\,K; G14) were used to determine that of the faint star. A value of $\Teff = \er{3377}{743}{199}$\,K was found; its large and asymmetric uncertainties are due to the large uncertainties in the measured values of $\Delta i$ and $\Delta z$, which are logarithmic quantities.

Magnitude differences and thus the fractional contributions of the faint companion to the total light of the system were then determined for the passbands of the light curves from Paper\,I. This was done using the same set of theoretical spectra, the Bessell $R$ and $I$ filter response functions from \citet{BessellMurphy12pasp} and the GROND $griz$ response functions. The results of this process are given in Table\,\ref{tab:lr}.

\reff{During the refereeing process of the current paper, magnitude differences in the $J$, $H$ and $K$ bands were published by \citet{Ngo+16apj} along with an improved separation measurement of $0.2397 \pm 0.0015$\,arcsec. These were obtained from adaptive-optics imaging with Keck/NIRC2 and are much more precise than the measurements of \citet{WollertBrandner15aa}: $\Delta J = 2.427 \pm 0.030$, $\Delta H = 2.217 \pm 0.010$ and $\Delta K_s = 1.965 \pm 0.019$. Although they require more extrapolation to visible wavelengths than the $\Delta i$ and $\Delta z$ measurements from \citet{WollertBrandner15aa}, their much greater precision means they provide better constraints on the properties of the companion star and thus on the effect of contaminating light on the transmission spectrum of WASP-103\,b. We have therefore repeated our analysis in order to include the new observations, and discuss the results from both analyses in the following work.}

\reff{The use of $\Delta J$ and $\Delta K_s$ versus $\Delta i$ and $\Delta z$ makes a significant difference to the contaminating light values and allows a major decrease in the uncertainties (Table\,\ref{tab:lr}). The change is particularly pronounced for the temperature of the faint star, which we now find to be $\Teff = \er{4405}{85}{80}$\,K, in agreement with the values found by \citet{Ngo+16apj}. Our results were obtained using the NIRC2 $J$ and $K_s$ transmission functions\footnote{{\tt http://www2.keck.hawaii.edu/inst/nirc2/\\filters.html}}; we checked the effect of using the 2MASS $J$ and $K_s$ transmission functions\footnote{{\tt http://www.ipac.caltech.edu/2mass/releases/\\allsky/doc/sec3\_1b1.html}} instead and found differences of no more than 0.25$\sigma$. This suggests that any imperfections in the available filter transmission functions have an insigificant effect on our results.}

\subsection{Light curve modelling}

\begin{figure} \includegraphics[width=\columnwidth,angle=0]{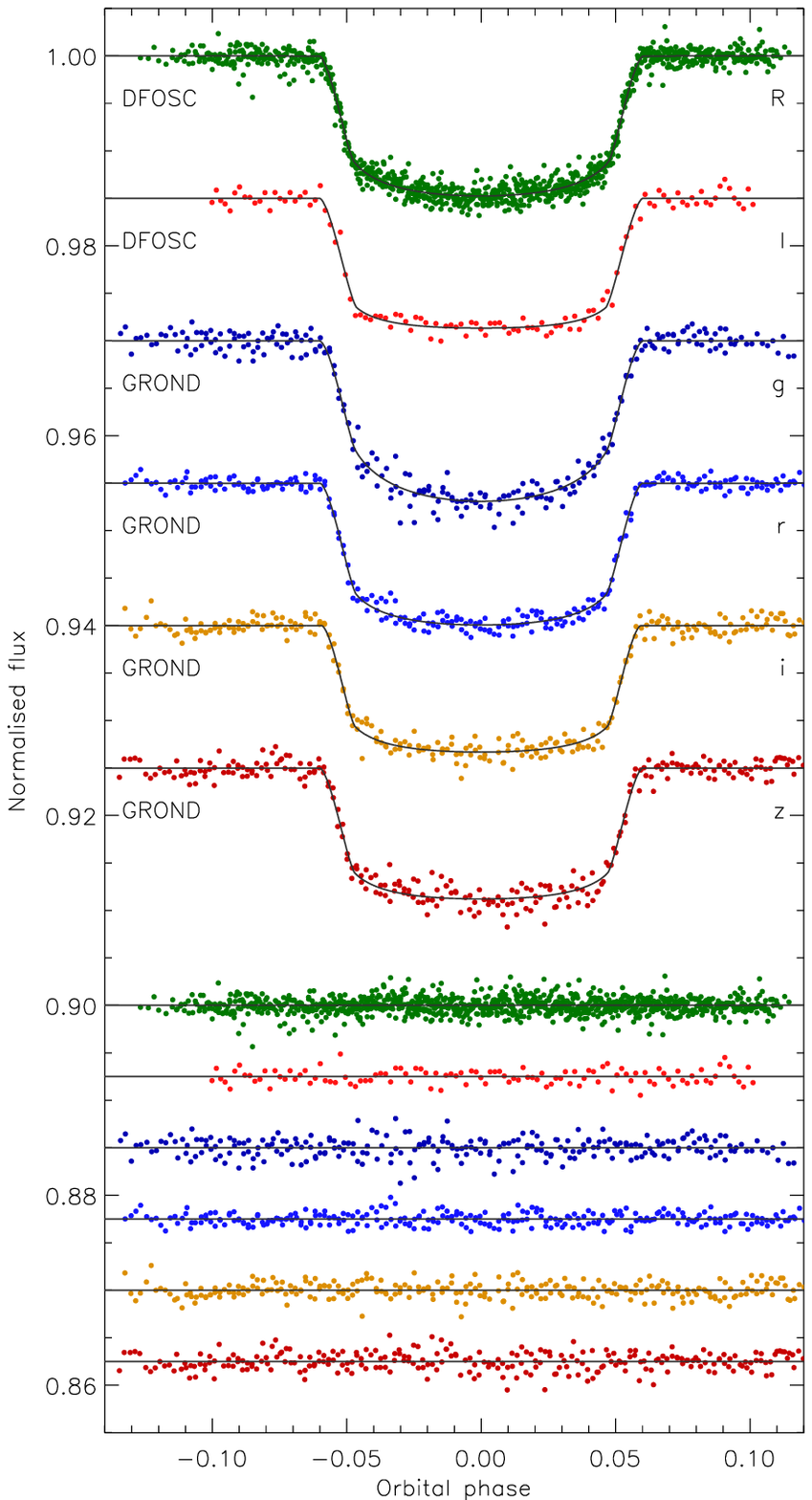}
\caption{\label{fig:lcfit} Phased light curves of WASP-103 from Paper\,I,
compared to the {\sc jktebop} best fits from the current work. The residuals
of the fits are plotted at the base of the figure, offset from unity. Labels
give the source and passband for each dataset. The polynomial baseline
functions have been removed from the data before plotting.} \end{figure}

The transit light curves were modelled in the same way as in Paper\,I, by using the {\sc jktebop} code \citep{Me13aa} to fit the data for each passband individually. The fitted parameters were the orbital inclination ($i$), reference time of mid-transit ($T_0$), and the sum and ratio of the fractional radii ($r_{\rm A} + r_{\rm b}$ and $k = \frac{r_{\rm b}}{r_{\rm A}}$) where the fractional radii are those of the star and planet in units of the orbital semimajor axis ($r_{\rm A,b} = \frac{R_{\rm A,b}}{a}$). Limb darkening was accounted for using four two-parameter laws, with the linear coefficient fitted and the non-linear coefficient fixed to theoretically-predicted values. Polynomials of order 1 or 2 versus time were used to normalise the flux to unity outside transit \citep{Me+14mn}. A circular orbit was assumed (G14). The one modification to the procedure was to include contaminating light (`third light' in {\sc jktebop} parlance) as a fitted parameter but constrained by the measured values collected in Table\,\ref{tab:lr}. Error estimates were obtained using Monte Carlo and residual-permutation simulations \citep{Me08mn} and the larger of the two options retained for each fitted parameter. The errorbars were further inflated to account for any scatter in the solutions for different limb darkening laws. The final results for each set of light curves are given in Table\,\ref{tab:lcfit} and the best fits are shown in Fig.\,\ref{fig:lcfit}.

\begin{table*} \centering \caption{\label{tab:lcfit} Parameters of the fit to the
light curves of WASP-103 from the {\sc jktebop} analysis (top). The final parameters
are given in bold and the parameters found by G14 and Paper\,I are given below this.}
\begin{tabular}{l r@{\,$\pm$\,}l r@{\,$\pm$\,}l r@{\,$\pm$\,}l r@{\,$\pm$\,}l r@{\,$\pm$\,}l}
\hline
Source              & \mc{$r_{\rm A}+r_{\rm b}$}   & \mc{$k$}                     & \mc{$i$ ($^\circ$)}  & \mc{$r_{\rm A}$}             & \mc{$r_{\rm b}$}                \\
\hline
DFOSC $R$-band      & \erc{0.370}{0.005}{0.002} & \erc{0.1160}{0.0010}{0.0006} & \erc{89.5}{1.5}{2.7} & \erc{0.331}{0.004}{0.002} & \erc{0.0384}{0.0006}{0.0002} \\
DFOSC $I$-band      & \erc{0.374}{0.016}{0.010} & \erc{0.1157}{0.0012}{0.0012} & \erc{85.6}{4.4}{3.3} & \erc{0.336}{0.014}{0.009} & \erc{0.0388}{0.0019}{0.0013} \\
GROND $g$-band      & \erc{0.372}{0.017}{0.007} & \erc{0.1196}{0.0021}{0.0016} & \erc{87.0}{3.0}{5.0} & \erc{0.332}{0.014}{0.006} & \erc{0.0397}{0.0023}{0.0010} \\
GROND $r$-band      & \erc{0.372}{0.011}{0.007} & \erc{0.1177}{0.0010}{0.0009} & \erc{86.3}{3.6}{3.0} & \erc{0.333}{0.009}{0.006} & \erc{0.0392}{0.0013}{0.0009} \\
GROND $i$-band      & \erc{0.364}{0.011}{0.003} & \erc{0.1126}{0.0015}{0.0014} & \erc{89.4}{0.6}{4.2} & \erc{0.327}{0.010}{0.003} & \erc{0.0368}{0.0014}{0.0006} \\
GROND $z$-band      & \erc{0.368}{0.012}{0.004} & \erc{0.1153}{0.0016}{0.0015} & \erc{89.9}{0.1}{4.4} & \erc{0.330}{0.010}{0.003} & \erc{0.0380}{0.0014}{0.0008} \\[3pt]
{\bf Final results} & \ercbf{0.3705}{0.0032}{0.0021} & \ercbf{0.1158}{0.0006}{0.0006} & \ercbf{88.2}{1.5}{1.5} & \ercbf{0.3319}{0.0030}{0.0019} & \ercbf{0.03854}{0.00041}{0.00030} \\[3pt]
Paper\,I            & 0.3712 & 0.0040              & 0.1127 & 0.0009              & 87.3 & 1.2           & 0.3335 & 0.0035              & 0.03754 & 0.00049               \\
G14                 & \mc{ }                       & \erc{0.1093}{0.0019}{0.0017} & 86.3 & 2.7           & \erc{0.3358}{0.0111}{0.0055} & \mc{0.03670}                    \\
\hline \end{tabular} \end{table*}

It can be seen that imposition of the contaminating light has led to parameter values which are generally consistent with previous results, \reff{and in some cases more precise due to the greater agreement between the models for different passbands}. The ratio of the radii \reff{is the exception as it has increased significantly (by 2.9$\sigma$),} as expected when third light is taken into account \citep[e.g.][]{Daemgen+09aa}.

%%%%%%%%%%%%%%%%%%%%%%%%%%%%%%%%%%%%%%%%%%%%%%%%%%%%%%%%%%%%%%%%%%%%%%%%%%%%%%%%%%%%%%%%%%%%%%%%%%%%%%%%%%%%%%%%%%%%%%%%%%%%%%%%%%%%%%%%%%%%%%%%%%%%%

\section{Physical properties of WASP-103}                                                                                          \label{sec:absdim}

\begin{table*} \caption{\label{tab:model} Derived physical properties of WASP-103. Quantities marked with a $^\star$ are significantly
affected by the spherical approximation used to model the light curves, and revised values are given at the base of the table.}
\begin{tabular}{l l l r@{\,$\pm$\,}c@{\,$\pm$\,}l r@{\,$\pm$\,}c@{\,$\pm$\,}l r@{\,$\pm$\,}l} \hline
Quantity                & Symbol           & Unit  & \mcc{\reff{This work}}                              & \mcc{Paper\,I}                  & \mc{G14}                       \\
\hline
Stellar mass            & $M_{\rm A}$      & \Msun & \ermcc{1.205}{0.094}{0.117}{0.021}{0.015}           & 1.204    & 0.089    & 0.019     & \erc{1.220}{0.039}{0.036}      \\
Stellar radius          & $R_{\rm A}$      & \Rsun & \ermcc{1.413}{0.040}{0.048}{0.008}{0.006}           & 1.419    & 0.039    & 0.008     & \erc{1.436}{0.052}{0.031}      \\
Stellar surface gravity & $\log g_{\rm A}$ & \,cgs & \ermcc{4.219}{0.012}{0.016}{0.003}{0.002}           & 4.215    & 0.014    & 0.002     & \erc{4.22}{0.12}{0.05}         \\
Stellar density         & $\rho_{\rm A}$   & \psun & \ercc{0.428}{0.007}{0.011}                          & \mcc{$0.421 \pm 0.013$}         & \erc{0.414}{0.021}{0.039}      \\[2pt]
Planet mass             & $M_{\rm b}$      & \Mjup & \ermcc{1.47}{0.11}{0.13}{0.02}{0.01}                & 1.47     & 0.11     & 0.02      & 1.490 & 0.088                  \\
Planet radius$^\star$   & $R_{\rm b}$      & \Rjup & \ermcc{1.596}{0.044}{0.054}{0.009}{0.007}           & 1.554    & 0.044    & 0.008     & \erc{1.528}{0.073}{0.047}      \\
Planet surface gravity  & $g_{\rm b}$      & \mss  & \ercc{14.34}{ 0.83}{ 0.85}                          & \mcc{$15.12 \pm  0.93$}         & 15.7 & 1.4                     \\
Planet density$^\star$  & $\rho_{\rm b}$   & \pjup & \ermcc{0.339}{0.023}{0.023}{0.001}{0.002}           & 0.367    & 0.027    & 0.002     & \erc{0.415}{0.046}{0.053}      \\[2pt]
Equilibrium temperature & \Teq\            & K     & \ercc{2489}{  66}{  65}                             & \mcc{$2495 \pm   66$}           & \erc{2508}{75}{70}             \\
Safronov number         & \safronov\       &       & \ermcc{0.0303}{0.0020}{0.0019}{0.0001}{0.0002}      & 0.0311   & 0.0019   & 0.0002    & \mc{ }                         \\
Orbital semimajor axis  & $a$              & au    & \ermcc{0.01979}{0.00051}{0.00065}{0.00012}{0.00008} & 0.01978  & 0.00049  & 0.00010   & 0.01985 & 0.00021              \\
Age                     & $\tau$           & Gyr   & \ermcc{3.8}{2.1}{1.7}{0.4}{0.4}                     & \ermcc{3.8}{2.1}{1.6}{0.3}{0.4} & \mc{3 to 5}                    \\
\hline
\multicolumn{6}{l}{{\it Planetary parameters corrected for asphericity:}} \\
Planet radius           &                  & \Rjup & \ercc{1.646}{0.052}{0.060}                          & \mcc{$1.603 \pm 0.052$} \\
Planet density          &                  & \pjup & \ercc{0.309}{0.025}{0.025}                          & \mcc{$0.335 \pm 0.025$} \\
\hline \end{tabular} \end{table*}

The physical properties of the WASP-103 system were measured in the same way as in Paper\,I (see also \citealt{Me12mn} and references therein) so we only briefly summarise the steps taken. The light curve parameters ($r_{\rm A}$, $r_{\rm b}$ and $i$) from Section\,\ref{sec:lc} were combined with the orbital period from Paper\,I and the spectroscopic parameters (\Teff, metallicity \FeH, and velocity amplitude $K_{\rm A}$) from G14. Five sets of tabulated predictions of theoretical stellar evolutionary models \citep{Claret04aa,Demarque+04apjs,Pietrinferni+04apj,Vandenberg++06apjs,Dotter+08apjs} were added. The overall best fit was then found to all properties using contraints from each of the theoretical models. Statistical errors were propagated from all input parameters, and systematic errors obtained from the scatter between the results for the five sets of theoretical models.

The final measurements for the physical properties of WASP-103 are given in Table\,\ref{tab:model}, which also hosts the values from Paper\,I and G14 for comparison. The only parameters for which the changes are worth mentioning are the radius \reff{(0.6$\sigma$)} and density \reff{(0.8$\sigma$)} of the planet, as expected. These are also the two parameters which need to be adjusted to account for the aspherical shape of the planet (see Paper\,I and \citealt{Budaj11aj}), and the corrections were applied in the same way as for Paper\,I. We can therefore conclude that the effect of the faint nearby star on the light curves of WASP-103 is insufficient to cause a significant change in the measured properties of the system. We next turn to the transmission spectrum, for which this statement certainly does not apply.

%%%%%%%%%%%%%%%%%%%%%%%%%%%%%%%%%%%%%%%%%%%%%%%%%%%%%%%%%%%%%%%%%%%%%%%%%%%%%%%%%%%%%%%%%%%%%%%%%%%%%%%%%%%%%%%%%%%%%%%%%%%%%%%%%%%%%%%%%%%%%%%%%%%%%

\section{The optical transmission spectrum of WASP-103\,b}

\begin{table} \centering
\caption{\label{tab:rb} Values of $r_{\rm b}$ for each of the light curves as
plotted in Fig.\,\ref{fig:rvary}. Note that the errorbars in this table exclude
all common sources of uncertainty in $r_{\rm b}$ so should only be used to compare
different values of $r_{\rm b}$ as a function of wavelength. The central wavelengths
and full widths at half maximum transmission are given for the filters used.}
\begin{tabular}{lccc} \hline
Passband & Central         & FWHM  & $r_{\rm b}$   \\
         & wavelength (nm) & (nm)  &               \\
\hline
$R$      & 658.9           & 164.7 & \er{0.03846}{0.00012}{0.00011} \\
$I$      & 820.0           & 140.0 & \er{0.03813}{0.00023}{0.00027} \\
$g$      & 477.0           & 137.9 & \er{0.03932}{0.00031}{0.00033} \\
$r$      & 623.1           & 138.2 & \er{0.03880}{0.00020}{0.00021} \\
$i$      & 762.5           & 153.5 & \er{0.03742}{0.00029}{0.00028} \\
$z$      & 913.4           & 137.0 & \er{0.03824}{0.00031}{0.00031} \\
\hline \end{tabular} \end{table}

\begin{figure} \includegraphics[width=\columnwidth,angle=0]{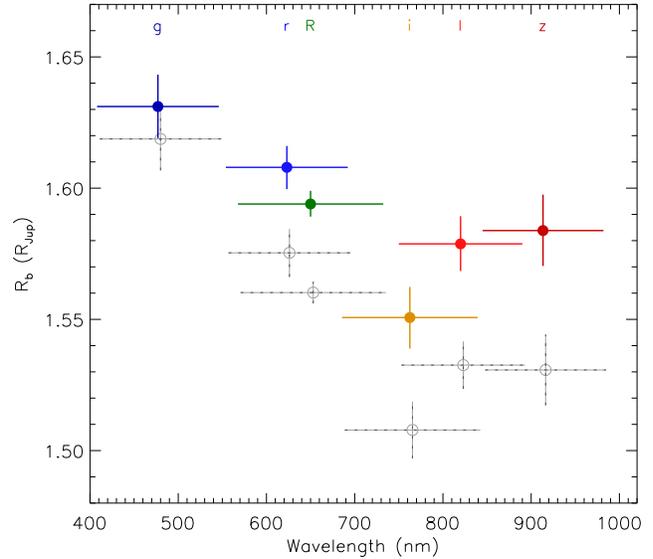}
\caption{\label{fig:rvary} Measured planetary radius ($R_{\rm b}$) as a function of the
central wavelength of the passbands used. Coloured and filled circles show the results
from this study, and the passbands are labelled at the top of the figure. The results from
Paper\,I are shown using grey open circles which have been offset by $+3$\,nm to bring
them out from underneath the newer results.} \end{figure}

\begin{figure} \includegraphics[width=\columnwidth,angle=0]{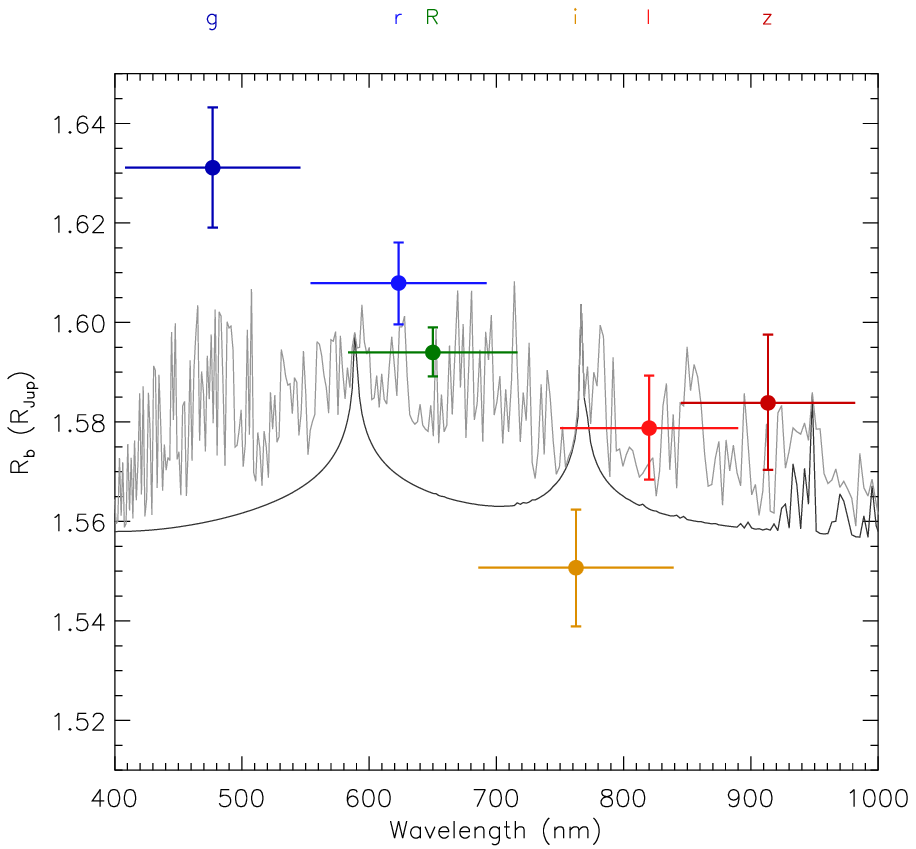}
\caption{\label{fig:nikku} Measured planetary radius ($R_{\rm b}$) as a function of
wavelength compared to the predictions from theoretical models of planetary atmospheres.
Filled circles show the results from this study, and the passbands are labelled at the
top of the figure. Transmission spectra are shown using grey lines and are for a planet
without TiO (dark grey smooth line) and with TiO (light grey jagged line). Both spectra
have been scaled to match the surface gravity of the planet and subsequently offset to
appear approximately in the centre of the plot.} \end{figure}

Following the approach of \citet{Me+12mn2}, we modelled the light curves of WASP-103 with the geometrical properties fixed to their best-fitting value from Table\,\ref{tab:lcfit}. The fractional radius of the planet and the linear LD coefficients were retained as fitted parameters, and the contaminating light was included as in Section\,\ref{sec:lc}. The data for each passband were fitted individually, and the true radius of the planet in that band found by multiplying $r_{\rm b}$ by the best-fitting value of $a$ ($41.41$\Rjup) neglecting its uncertainty. This process yielded measurements of the radius of the planet in different passbands, calculated in a consistent way and ignoring sources of error common to all passbands (e.g.\ the uncertainty in $a$). The results can be found in Table\,\ref{tab:rb} along with the characteristics of the filters used for the observations.

Fig.\,\ref{fig:rvary} shows the measured planetary radius as a function of wavelength. The results from Paper\,I are also shown for comparison. \reff{Significant differences are seen in all passbands except the $g$-band}. It is clear that the inclusion of the faint star in the analysis changes the interpretation of the optical transmission spectrum of WASP-103\,b\reff{, primarily by weakening the variation of radius with wavelength.}

In Fig.\,\ref{fig:nikku} we compare the radius measurements of WASP-103\,b to two representative theoretical transmission spectra. The spectra are for planets with and without titanium oxide \citep{Hubeny++03apj,Fortney+08apj} and were kindly calculated by Nikku Madhusudhan (see \citealt{MadhusudhanSeager09apj} and \citealt{MadhusudhanSeager10apj}). They have been scaled to the surface gravity and temperature of WASP-103\,b, and then arbitrarily offset to appear near the centre of the plot. It is clear that neither match the observations well, and in particular predict a rise in radius near the centre of the plot whereas the observations themselves show the opposite. The same conclusion is reached when considering alternative sets of model transmission spectra \citep{Fortney+08apj,Fortney+10apj}.

\subsection{The view from MassSpec}

\begin{figure} \includegraphics[width=\columnwidth,angle=0]{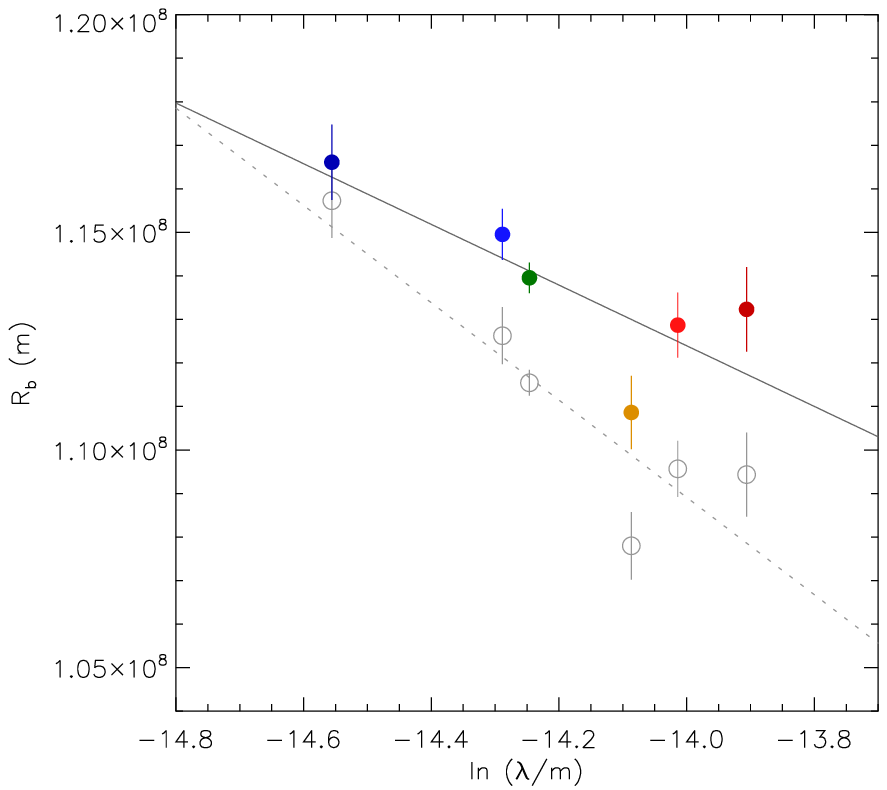}
\caption{\label{fig:massspec} Variation of planet radius with wavelength in SI units
(filled circles) and the slope $-\frac{{\rm d}R_{\rm b}(\lambda)}{{\rm d}\ln\lambda}$
determined from the data (unbroken line). For comparison the values from Paper\,I are
shown with grey open circles and a grey dotted line.} \end{figure}

The MassSpec concept \citep{DewitSeager13sci} is that it is possible to measure the mass of a planet from its transmission spectrum. It depends on obtaining the atmospheric scale height, $H$, from the power-law variation of planet radius with wavelength:
\begin{equation}
\alpha H = \frac{{\rm d}R_{\rm b}(\lambda)}{{\rm d}\ln\lambda}
\end{equation}
where $\alpha = 4$ for Rayleigh scattering \citep{Lecavelier+08aa} and can take other values for different scattering processes. The atmospheric scale height depends on surface gravity and therefore the mass of the planet:
\begin{equation}
H = \frac{ k_{\rm B} \Teq }{ \mu \, g_{\rm b} } = \frac{ k_{\rm B} T_{\rm b} R_{\rm b}^{\,2} }{ \mu G M_{\rm b} }
\end{equation}
where $T_{\rm b}$ is the local temperature, $\mu$ is the mean molecular weight of the atmosphere (assumed to be 2.3\,a.m.u.\ for gas giants), G is the Newtonian gravitational constant \reff{and $k_{\rm B}$ is Boltzmann's constant}. We recast this equation to include all measurable terms on the left and all quantities which might be sought on the right (contrast with eq.\,4 in Paper\,I):
\begin{equation}
-\frac{{\rm d}R_{\rm b}(\lambda)}{{\rm d}\ln\lambda} \frac{G}{k_{\rm B}R_{\rm b}^{\,2}} = \frac{ \alpha k_{\rm B} T_{\rm b} }{ \mu M_{\rm b} } \label{eq:big}
\end{equation}

In Paper\,I we used this approach to infer a mass of $0.31 \pm 0.05$\Mjup\ for the planet, with a significance of $7.3\sigma$, which was much lower than the dynamically-measured value (Table\,\ref{tab:model}). Our new analysis returns a \reff{higher} mass of $0.53 \pm 0.13$\Mjup, with the variation of radius with wavelength ($-\frac{{\rm d}R_{\rm b}(\lambda)}{{\rm d}\ln\lambda}$) measured to a \reff{lower} significance level of $4.4\sigma$ based on a simple Monte Carlo propagation of the uncertainties (see Fig.\,\ref{fig:massspec}). In both cases we have adopted $\mu = 2.3$, $\alpha = 4$ and $T_{\rm b} = \Teq$. \reff{This mass is still far too low to match the value found from the analysis in Section\,\ref{sec:absdim}.}

However, it is possible to adjust any of the quantities on the right-hand side of Eq.\,\ref{eq:big} to match the measured value of the left-hand side. It is therefore reasonable to seek the value of $\alpha T_{\rm b}$ where the planet mass from MassSpec equals the dynamical measurement \citep[e.g.][]{Sing+11mn,Nikolov+15mn}. \reff{By manual iteration we found $\alpha T_{\rm b} = 27900 \pm 2200$, and therefore $\alpha = 11.2 \pm 0.9$ when using $T_{\rm b} = \Teq$, which is extremely high. Alternatively, a value for the mean molecular weight of $\mu = 0.83 \pm 0.07$\,a.m.u.} would balance the equation but is unphysically small. We therefore conclude that the slope of radius versus wavelength measured for WASP-103\,b remains too large to be explained by Rayleigh scattering, as expressed in the MassSpec paradigm.

%%%%%%%%%%%%%%%%%%%%%%%%%%%%%%%%%%%%%%%%%%%%%%%%%%%%%%%%%%%%%%%%%%%%%%%%%%%%%%%%%%%%%%%%%%%%%%%%%%%%%%%%%%%%%%%%%%%%%%%%%%%%%%%%%%%%%%%%%%%%%%%%%%%%%

\section{Physical properties of the companion and hierarchy of the system}                                                           \label{sec:comp}

\citet{WollertBrandner15aa} did not attempt to characterise the companion, or assess the chance that it is bound to the planetary system. We therefore now estimate its mass, the probability that it is a foreground or background star, and the probability that it is physically bound given the binary frequency in the solar neighbourhood.

The only observed quantities available for the nearby companion star are its sky position and magnitude differences in \reff{the $i$, $z$, $J$, $H$ and $K_s$ filters} versus the planet host star. These magnitude differences were found to correspond to ${\Teff}_{\rm,comp} = \er{4405}{85}{80}$\,K in Section\,\ref{sec:lc}. Using the temperature--mass calibration presented by \citet{Evans+16aa} we find a \reff{mass of $M_{\rm comp} = 0.72 \pm 0.08$\Msun, where} the errorbars include an astrophysical scatter of 0.08\,dex in $\log M_{\rm comp}$ added in quadrature to the error arising from the uncertainty in ${\Teff}_{\rm,comp}$. \reff{This is in good agreement with the value of $M_{\rm comp} = 0.721 \pm 0.024$\Msun\ derived by \citet{Ngo+16apj}. The companion star is therefore probably a mid-to-late K dwarf.}

The probability that the two stars form an asterism was calculated as follows. We used the TRILEGAL galactic model \citep{Girardi+05aa} to produce a synthetic population of stars for a $1^\circ$ field centred on WASP-103 ($l=23.4^\circ,\ b=+33.0^\circ$). Stars were simulated to a depth of $i=26$, and the default parameters for version 1.6 of the model were used \citep[see e.g.][]{LilloBox++14aa,Evans+16aa}.

\reff{The 2MASS catalogue \citep{Cutri+03book} does not resolve the two stars, giving a combined value of $K_s = 10.767 \pm 0.020$, which we assume includes all light from both WASP-103\,A and the companion. \citet{Ngo+16apj} measured a $K_s$ magnitude difference of $1.965 \pm 0.019$, from which we calculate apparent magnitudes $Ks = 10.932 \pm 0.020$ for WASP-103 and $Ks = 12.897 \pm 0.026$ for the companion. We note that \citet{Ngo+16apj} also calculated the apparent magnitudes and colours for their detected companions based on 2MASS data, but under the assumption that the 2MASS magnitude represents flux from only the planet-host star. In the case of WASP-103 this assumption is not valid, because the companion is completely unresolved in the 2MASS data and contributes $16\%$ of the total flux.}

We binned our synthetic population by $K_s$ magnitude, weighted the bins by the probability of the companion having the corresponding $K_s$ magnitude, and then summed over all bins to determine the density of stars with the companion's magnitude. The weighted stellar density was then multiplied by the sky area contained by a \reff{circle with radius $0.2397\as$, to give a probability of $2.6\times10^{-9}$ that a star of this $K_s$ magnitude is present within $0.2397\as$ of the planet host star}.

We then checked the probability that the companion star is physically bound to WASP-103. To compare this hypothesis to the unbound scenario, we determined the fraction of stellar systems we would expect to have the measured magnitude difference and projected separation using a Monte Carlo simulation. We adopted the binary population model used in \citet{Evans+16aa}, which utilises the population data from \citet{Raghavan+10apjs}. 46\% of stars were assumed to be in binaries, period (in days) was log-normally distributed with a mean of 5.03 and a standard deviation of 2.28, eccentricity was uniformly distributed in the interval [0.0,0.95], and the mass ratio followed a three-part parameterisation to represent the high frequency of mass ratios near $1.0$ and low fraction of binaries with extreme mass ratios. All other orbital elements were randomised, including the phase at the time of observation. Projected separations were calculated assuming a distance of $470 \pm 35$\,pc (G14), and the magnitude difference determined from the models of \citet{An+09apj}, with $\FeH = 0.06$ and an age of 3.6\,Gyr. The simulated stars were then weighted based on how closely they matched the measured magnitude difference of the companion, from which we determined that the probability of a star similar to WASP-103 having a matching stellar companion within $0.2397\as$ is $0.0013$. This is \reff{five} orders of magnitude more likely than the background star scenario, even though the inclusion of a constraint on mass ratio makes the probability conservatively low, so we conclude that the faint companion forms a part of the WASP-103 system.

Assuming the angular separation and distance to the system as above, the projected separation of the two stars is $113 \pm 8$\,AU. Its orbital period is therefore of order 1000\,yr, so is far too long for confirmation via spectroscopic radial velocity measurements. Additional flux ratios covering a wider wavelength range will, however, allow us to pin down its \Teff\ and therefore mass and radius much more precisely.

\begin{figure} \includegraphics[width=\columnwidth,angle=0]{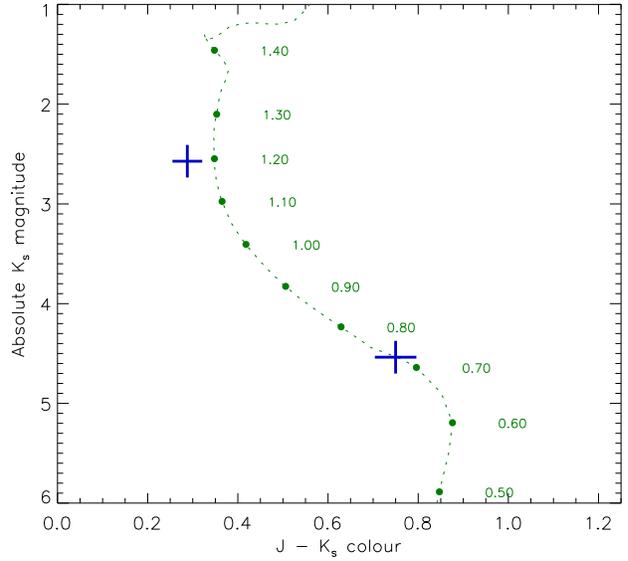}
\caption{\label{fig:iso} Isochrone from \citet{An+09apj} for an age of 3.6\,Gyr
and $\FeH = 0.06$ compared to \reff{the $J-K_s$ colour and $K_s$-band} absolute magnitude
of the two stars. The isochrone is plotted with a green dotted line and labelled
with the initial stellar mass at appropriate intervals. The properties of the
two stars are shown with thick blue lines.} \end{figure}

In Fig.\,\ref{fig:iso} we plot the absolute magnitudes of the two stars against the theoretical isochrone from \citet{An+09apj} for an age of 3.6\,Gyr and $\FeH = 0.06$. The position of the planet host star is in excellent agreement with its measured mass, \reff{confirming the correctness of the distance measurement given by G14. The $K_s$-band magnitude difference suggests a mass of around 0.7\Msun\ for the secondary star, in agreement with \citet{Ngo+16apj} despite the use of different synthetic spectra ({\sc atlas9} versus {\sc phoenix}) and theoretical evolutionary models (\citealt{An+09apj} versus \citealt{Baraffe+98aa}).}

%%%%%%%%%%%%%%%%%%%%%%%%%%%%%%%%%%%%%%%%%%%%%%%%%%%%%%%%%%%%%%%%%%%%%%%%%%%%%%%%%%%%%%%%%%%%%%%%%%%%%%%%%%%%%%%%%%%%%%%%%%%%%%%%%%%%%%%%%%%%%%%%%%%%%

\section{Summary and discussion}                                                                                                     \label{sec:disc}

The WASP-103 planetary system is an important tracer of the atmospheric properties of very hot planets, as its short orbital period and hot host star lead to a high equilibrium temperature of $2489 \pm 66$\,K. This is even hotter than that for WASP-121\,b \citep{Delrez+16mn}, which has very recently been found to have titanium oxide and vanadium oxide absorption in its atmosphere based on spectroscopy from the Hubble Space Telescope \citep{Evans+16apjl}, although we note that thermal inversions can arise due to other chemical species \citep{Molliere+15apj}. The first work to probe the atmosphere of WASP-103\,b (Paper\,I) found that the radius of the planet -- as measured by transit depth -- was greater at bluer optical wavelengths. The significance level of this detection was strong at $7.3\sigma$, but the slope was much greater than expected from Rayleigh scattering and therefore its physical interpretation was not clear.

The subsequent detection of a faint nearby star \citep{WollertBrandner15aa} offered the possibility of removing the discrepancy, by explaining the radius variation as a result of light from a faint and red object contaminating the light curves of WASP-103, rather than an intrinsic property of the planetary system. \reff{We have therefore used $J$- and $K$-band magnitude differences between the planet host and the nearby star, recently presented by \citet{Ngo+16apj},} to determine the amount of contaminating light as a function of passband, apply the corrections in a reanalysis of the transit light curves, and rederive the properties of the system plus the optical transmission spectrum of the planet.

\reff{The effect of the inclusion of contaminating light is significant on some the photometric parameters, in particular the ratio of the radii, which has increased by 2.9$\sigma$. This quantity is typically the best-determined of the photometric parameters because it depends directly on the transit depth and is only weakly correlated with other parameters. It is also generally found to be the quantity which exhibits the worst agreement between different datasts (e.g.\ \citealt{Me09mn,Me10mn,Me11mn,Me12mn}). We have usually ascribed this issue to the very fact that the ratio of the radii is the best-determined photometric parameter, so therefore is the parameter which is most sensitive to the existence of red noise in transit light curves. It is also clear that some of this discord can be attributed to the variation of opacity with wavelength, which is the underlying physical process probed using transmission spectroscopy and photometry. However, it is likely that some is due to flux contributions arising from undetected faint nearby stars, in which case the scatter in the ratio of the radii is not intrinsic to many of the planetary systems which have been studied in the past.}

The effect of the inclusion of contaminating light on the measured physical properties, however, is somewhat smaller. The main changes for the WASP-103 system are that the measured planet radius increases \reff{by 0.6$\sigma$ and the density decreases by 0.8$\sigma$}. This is encouraging in that our understanding of the general planet population is not greatly affected by the presence of undetected close companions (see also \citealt{Daemgen+09aa}). Such a statement is not valid in general, however, as light curves in redder bands are more affected in the typical scenario whereby the faint star is redder than the planet host star. Whilst there are many advantages in observing at redder optical wavelengths, such as weaker stellar limb darkening and starspot perturbations, it is better to observe in bluer passbands if contaminating light from a redder star is an issue. This will be an important consideration for the TESS mission \citep{Ricker+14spie}, which has a very coarse pixel scale and a passband which cuts on at 600\,nm in order to minimise the chromatic aberrations present in refractive optics.

The impact of the contaminating light on the transmission spectrum of the planet is much more important. Instead of a relatively featureless slope throught the optical wavelength range, there is now an upward inflection redward of 760\,nm. This is not reproduced by existing theoretical models of planetary transmission spectra, which tend to predict the opposite: larger radii in the middle of the optical band due to broad absorption from either atomic sodium and potassium, or titanium oxide. An explanation of the transmission spectrum of WASP-103\,b demands strong absorption from species at both bluer and redder wavelengths.

The MassSpec concept was invoked to explore possible explanations of the transmission spectrum of WASP-103\,b. We note that this is not strictly applicable, because the spectrum does not exhibit a monotonic slope through the optical wavelength range and therefore is not consistent with purely scattering processes. The spectral slope is much weaker once the contaminating light has been included in the analysis, and is also more uncertain due to the large errorbars in the measurements of the contaminating light. \reff{The slope now has a significance of $4.4\sigma$, and corresponds to a planet mass of $0.53 \pm 0.13$\Mjup\ which is still much smaller than the dynamically-measured value of $1.47 \pm 0.11$\Mjup.}

\reff{The magnitude differences $\Delta J$ and $\Delta K_s$ between the companion and the planet host star are consistent with the companion having a temperature of ${\Teff}_{\rm,comp} = \er{4405}{85}{80}$\,K and thus a mass of $M_{\rm comp} = 0.72 \pm 0.08$\Msun. The probability of the two stars being aligned by chance is very low, $2.6\times10^{-9}$,} so they are almost certainly gravitationally bound. WASP-103 is therefore a hierarchical system consisting of (at least) two stars and one planet.

It is clear that our understanding of the WASP-103 system remains incomplete. A major improvement could be obtained from more precise characterisation of the flux ratio between the planet host and the faint nearby star. \reff{The available flux ratios are either extremely uncertain \citep{WollertBrandner15aa} or require extrapolation from near-infrared to optical wavelengths \citep{Ngo+16apj}.} Additional observations should be obtained \reff{at optical wavelengths} using adaptive optics on a large telescope, which can be capable of much greater resolution than a lucky imager on a 2.2\,m telescope. It is entirely possible that such observations will cause a further revision to the measured transmission spectrum of WASP-103\,b.

Finally, the WASP-103 system is now known to be another example of a planet in a stellar binary system. It is therefore an important tracer of the formation mechanisms of binary and planetary systems \citep{Desidera+14aa,Neveu+14aa,Neveu+16aa}.

%%%%%%%%%%%%%%%%%%%%%%%%%%%%%%%%%%%%%%%%%%%%%%%%%%%%%%%%%%%%%%%%%%%%%%%%%%%%%%%%%%%%%%%%%%%%%%%%%%%%%%%%%%%%%%%%%%%%%%%%%%%%%%%%%%%%%%%%%%%%%%%%%%%%%

\section*{Acknowledgements}

JS acknowledges financial support from the Leverhulme Trust in the form of a Philip Leverhulme Prize. DFE acknowledges financial support from STFC in the form of a PhD studentship. We thank Barry Smalley, Pierre Maxted and Luigi Mancini for discussions, and an anonymous referee for helpful comments. The following internet-based resources were used in research for this paper: the ESO Digitized Sky Survey; the NASA Astrophysics Data System; the SIMBAD database and VizieR catalogue access tool operated at CDS, Strasbourg, France; and the ar$\chi$iv scientific paper preprint service operated by Cornell University.

%%%%%%%%%%%%%%%%%%%%%%%%%%%%%%%%%%%%%%%%%%%%%%%%%%%%%%%%%%%%%%%%%%%%%%%%%%%%%%%%%%%%%%%%%%%%%%%%%%%%%%%%%%%%%%%%%%%%%%%%%%%%%%%%%%%%%%%%%%%%%%%%%%%%%

\bibliographystyle{mn_new}
% \bibliography{aamnem99,jkt}
% \bsp

%%%%%%%%%%%%%%%%%%%%%%%%%%%%%%%%%%%%%%%%%%%%%%%%%%%%%%%%%%%%%%%%%%%%%%%%%%%%%%%%%%%%%%%%%%%%%%%%%%%%%%%%%%%%%%%%%%%%%%%%%%%%%%%%%%%%%%%%%%%%%%%%%%%%%
\end{document}